\documentclass[
  reprint, 
  superscriptaddress, 
  amsmath,amssymb,
  aps, prb, 
  floatfix,
]{revtex4-2}

\usepackage{graphicx}
\usepackage{dcolumn}
\usepackage{bm}
\usepackage[mathlines]{lineno}
\usepackage[utf8]{inputenc}
\usepackage[T1]{fontenc}
\usepackage{xcolor}
\usepackage[colorlinks=true,
            linkcolor=blue,
            urlcolor=blue,
            citecolor=blue]{hyperref}

\begin{document}

\title{\texorpdfstring{$H$}{H}-linear magnetoresistance in the \texorpdfstring{$T^2$}{T2} resistivity regime of overdoped infinite-layer nickelate \texorpdfstring{La$_{1-x}$Sr$_x$NiO$_2$}{La1-xSrxNiO2}}

\author{Yong-Cheng Pan}
\affiliation{Department of Physics, National Tsing Hua University, Hsinchu 300044, Taiwan}

\author{Tommy Kotte}
\affiliation{Dresden High Magnetic Field Laboratory (HLD-EMFL), Helmholtz-Zentrum Dresden-Rossendorf, Dresden 01328, Germany}

\author{Toni Helm}
\affiliation{Dresden High Magnetic Field Laboratory (HLD-EMFL), Helmholtz-Zentrum Dresden-Rossendorf, Dresden 01328, Germany}

\author{Motoki Osada}
\affiliation{Quantum-Phase Electronics Center and Department of Applied Physics, The University of Tokyo, Tokyo 113-8656, Japan}

\author{Atsushi Tsukazaki}
\affiliation{Quantum-Phase Electronics Center and Department of Applied Physics, The University of Tokyo, Tokyo 113-8656, Japan}
\affiliation{Institute for Materials Research, Tohoku University, Sendai, Miyagi, Japan}

\author{Yu-Te Hsu}
\email{ythsu@phys.nthu.edu.tw}
\affiliation{Department of Physics, National Tsing Hua University, Hsinchu 300044, Taiwan}
\affiliation{Center for Science and Quantum Technology, National Tsing Hua University, Hsinchu 300044, Taiwan}
\affiliation{Department of Materials Science and Engineering, National Tsing Hua University, Hsinchu 300044, Taiwan}

\date{\today}

\begin{abstract}
We report a systematic magnetotransport study on high-crystallinity La$_{1-x}$Sr$_{x}$NiO$_2$ (LSNO) thin films with $x=0.20-0.24$. By conducting pulsed-field transport experiment up to 62~T, we reveal two salient features of the normal-state transport in overdoped LSNO thin films: (1) the magnetoresistance does not follow the Kohler's rule but exhibits a $H$-linear behavior in the high $H/T$ limit and (2) the normal-state $\rho(T)$ below 30~K consistently follows a $T^2$ behavior across the overdoped regime. Our results demonstrate a coexistence of $H$-linear magnetoresistance and $T^2$ resistivity in a model unconventional superconductor and provide new information on the transport characteristics of the normal ground state that host superconductivity in infinite-layer nickelates.
\end{abstract}

\maketitle


\section{Introduction}
In conventional metals with highly correlated electrons, the temperature-dependent electrical resistivity in the $T=0$ limit follows a quadratic power law i.e. $\Delta\rho(T)\propto T^2$, one of the defining signatures of a Fermi-liquid ground state \cite{mackenzie1996,maeno1997,nakamae2003,perry2006,proust2016}. Similarly, the magnetic-field-dependent resistivity follows a quadratic function form i.e. $\Delta\rho(H)\propto H^2$ as a result of carrier orbital motion in the presence of applied magnetic field \cite{pippard1989}. The magnetoresistance ($\mathrm{MR}=\Delta\rho({H},T)=\rho(H,T)-\rho(0,T)$), defined as the magnitude of the field-induced change in resistivity at a given temperature, is thus dictated by the carrier mean free path $\ell$, which is inversely proportional to the zero-field resistivity $\rho(0,T)$. 
In systems where the transport behavior is characterized by a single isotropic carrier lifetime $\tau$, the MR is determined by $\ell\sim\omega_{\rm c}\tau$, where $\omega_{\rm c}$ is the cyclotron frequency; consequently, MR normalized by the zero-field resistivity ($\Delta\rho(H,T)/\rho(0,T)$) at various temperatures collapse into a single curve when plotted against $\mu_0H/\rho(0)$, known as the Kohler's rule \cite{kohler1938,chan2014}. The exhibition of a $T^2$ resistivity as $T\rightarrow 0$ and a MR following Kohler's rule is considered strong evidence for charge transport described by the Fermi-liquid framework.

By contrast, in solid-state systems near a quantum critical point (QCP), the zero-temperature endpoint of a continuous phase transition, the Fermi-liquid description breaks down and the transport behavior often displays a remarkable deviation from the quadratic temperature and field dependence. For example, it is well established that a $T$-linear resistivity replaces the $T^2$ power law near the QCP\cite{loehneysen2007,shibauchi2014,licciardello2019}. 
Moreover, in several unconventional superconductors near their putative QCP \cite{hayes2016,licciardello2019a,giraldogallo2018,sakar2019,ayres2021,ayres2024}, the normal-state magnetoresistivity follows a distinct functional form: 
\textcolor{black}{
\begin{equation}
    \rho(H,T) = \mathcal F(T)+\sqrt{(\alpha T)^2 + (\gamma \mu_0H)^2},
\end{equation}
where the first term is associated with inelastic scatterings, the second term encapsulates the elastic (MR) response, and} $\alpha$ and $\gamma$ are numeric constants. The observation of the unconventional MR described by Eq.~(1) can be demonstrated by a $H$-linear MR with a constant slope as $T\rightarrow0$ and the collapse of $\Delta\rho(H,T)/T$ against $H/T$ onto a single curve \cite{hayes2016,ayres2021}. The concurrence of a $T$-linear $\rho(T)$ and $H$-linear MR have been recently been employed as a diagnostic tool for a \textit{strange metal} state, a remarkable breakdown of the Fermi-liquid description, 
\textcolor{black}{suggestive of a universal bound for the inelastic scattering rate set by Planck's constant \cite{bruin2013,legros2018,hartnoll2022,shekhter2025}.}

In infinite-layer nickelates, a consensus on the ground state description in the region where superconductivity is found has not been reached \cite{wang2024,wang2025,puphal2025}. A major obstacle lies in the challenging synthesis required for preparing infinite-layer nickelate thin films over a wide range of doping with consistently high crystallinity. Continuous efforts on improving the sample quality of nickelate thin films have led to notable recent advancements in the nickelate physics, including mapping of the Fermi surface in La$_{2-x}$(Sr/Ca)$_x$NiO$_2$ by angle-resolved photoemission spectroscopy \cite{ding2024,sun2025} and systematic studies of the normal-state transport in both the Nd$_{1-x}$Sr$_x$NiO$_2$ (NSNO) \cite{lee2023,hsu2024} and La$_{1-x}$Sr$_x$NiO$_2$ (LSNO) families \cite{osada2025}. As recent experiments have shown that the rare-earth magnetism plays a significant role in determining the normal-state and superconducting properties \cite{fowlie2022,wang2023,chow2022,harvey2025}, the La-based compound offers an advantage of probing the physics of infinite-layer nickelates without the complication of rare-earth magnetism \cite{osada2021,zeng2021b}. 

Intriguingly, several recent experimental studies have reported signatures associated with quantum criticality in superconducting LSNO films, including (1) a $T$-linear normal-state resistivity over an extended temperature range \cite{osada2023}, (2) a logarithmic divergence of Seebeck coefficient ($S/T\propto \mathrm{log}(1/T)$) below 60~K \cite{iorioDuval2025}; and (3) a concerted sign change in both Hall coefficient and Seebeck coefficient near the optimal doping $x_{\rm opt}=0.16$ \cite{osada2025}. While these findings hint at the existence of a possible QCP near $x_{\rm opt}$, a systematic study of the normal-state resistivity and magneotoresistance in the $T=0$ limit, leaving open the question of whether the normal ground state of LSNO is a \textit{bona fide} strange metal. 

In this work, we present a systematic magnetotransport study on high-crystallinity LSNO thin films in the overdoped regime ($x=0.20-0.24$). By conducting pulsed-field transport experiment up to 62~T, we uncover the normal-state transport behavior down to $\simeq$~2~K and find that the normal-state MR in LSNO does not follow the Kohler's rule but can be described using the functional form of Eq.~(1). Meanwhile, the normal-state $\rho(T)$ below 30~K is consistent with a $T^2$ description in all three films studied. Our results demonstrate an intriguing dichotomy between the inelastic and elastic channels of charge transport in a model correlated electron system.

\section{Methods}
LSNO films were synthesized using a pulsed-laser deposition technique\cite{osada2023}. In short, La$_{1-x}$Sr$_x$NiO$_3$ films of $\approx$ 9~nm thickness, capped with SrTiO$_3$ layer of 5 unit cells, were first grown on SrTiO$_3$ (001) substrates with a KrF excimer laser ($\lambda=248$~nm). The La$_{1-x}$Sr$_x$NiO$_3$ films were subsequently wrapped in aluminum foil, placed in a Pyrex glass tube with CaH$_2$ powder ($\sim$0.1~g), and annealed in a tube furnace at 340 to 370~$^\circ$C under vacuum (below 10~mtorr) for a total reduction time of 2.5 to 4 hours. Stabilization of the infinite-layer phase was confirmed by X-ray diffractions. 
Standard four-probe electrical contacts were made using indium leads, then extended to the measurement platform using gold wires and room-temperature silver paint (DuPont 4929N). 
For all reported results, the electrical current was applied within the crystalline $ab$-plane and magnetic field applied perpendicular to the sample plane i.e. parallel to the crystalline $c$-axis. Resistivity measurements in low magnetic fields ($<9$~T) were performed using the Quantum Design Physical Property Measurement System (PPMS) as well as the Electrical Transport Option of Magnetic Property Measurement System (MPMS3-ETO) with an excitation current amplitude of 10~$\mu$A and a frequency $\simeq20$~Hz.
Magnetotransport measurements in magnetic fields up to 62~T were performed in the High Field Laboratory (HLD-EMFL) in Dresden, Germany. Excitation currents of 20~$\mu$A amplitude with frequency $\simeq$ 1~kHz.

\section{Results}
\subsection{Zero-field resistivity}
\begin{figure}[h!!!]
\includegraphics[width=1\linewidth]{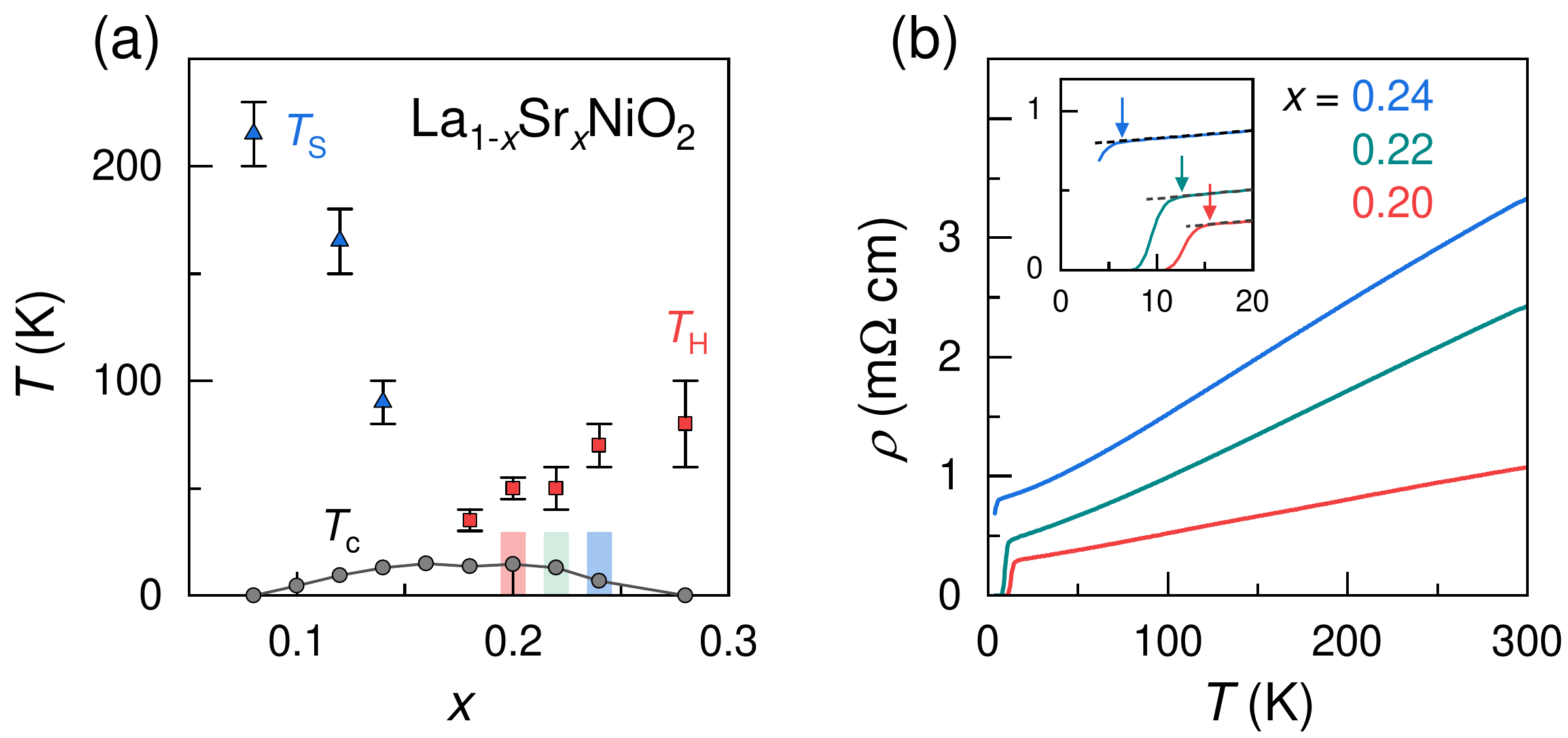}
\caption{Phase diagram of La$_{1-x}$Sr$_x$NiO$_2$ (LSNO) and zero-field resistivity of samples studied in this work.
(a) Characteristic temperatures of LSNO thin films as a function of Sr doping $x$. $T_{\rm c}$: superconducting critical temperature; $T_{\rm H}$: Hall-coefficient sign-change temperature; $T_{\rm S}$: Seebeck-coefficient sign-change temperature. 
Color shadings mark the doping levels studied in this work. $T_{\rm c}, T_{\rm H}$ and $T_{\rm S}$ reproduced from \cite{osada2025}.
\textcolor{black}{(b) In-plane resistivity versus temperature $\rho(T)$ measured at zero applied magnetic field for $x$ = 0.20, 0.22, and 0.24. 
$T_{\rm c}$ values are 15.6, 12.6, 6.4~K, respectively, for the $x=0.20$, 0.22 and 0.24 films (marked by the vertical arrows in the inset), corresponding to the onset temperature of a deviation from the linearly extrapolated normal-state behavior (gray dash lines).}}
\label{Fig_phaseDiagram}
\end{figure}

Figure~\ref{Fig_phaseDiagram}(a) shows the phase diagram of LSNO constructed using transport experiments as recently reported in \cite{osada2025}. The optimal doping $x_{\rm opt}$ is $\approx 0.16$, around which both the Seebeck and Hall coefficient shows a change of sign in the $T=0$ limit. This observation suggests a transition of the underlying electronic structure and correlations across $x_{\rm opt}$. In this study, we focus on the overdoped regime in the LSNO phase diagram within which the electron correlation is found to be diminished compared to the underdoped side \cite{osada2025}.

\begin{figure*}[hbtp!!!]
\includegraphics[width=0.8\linewidth]{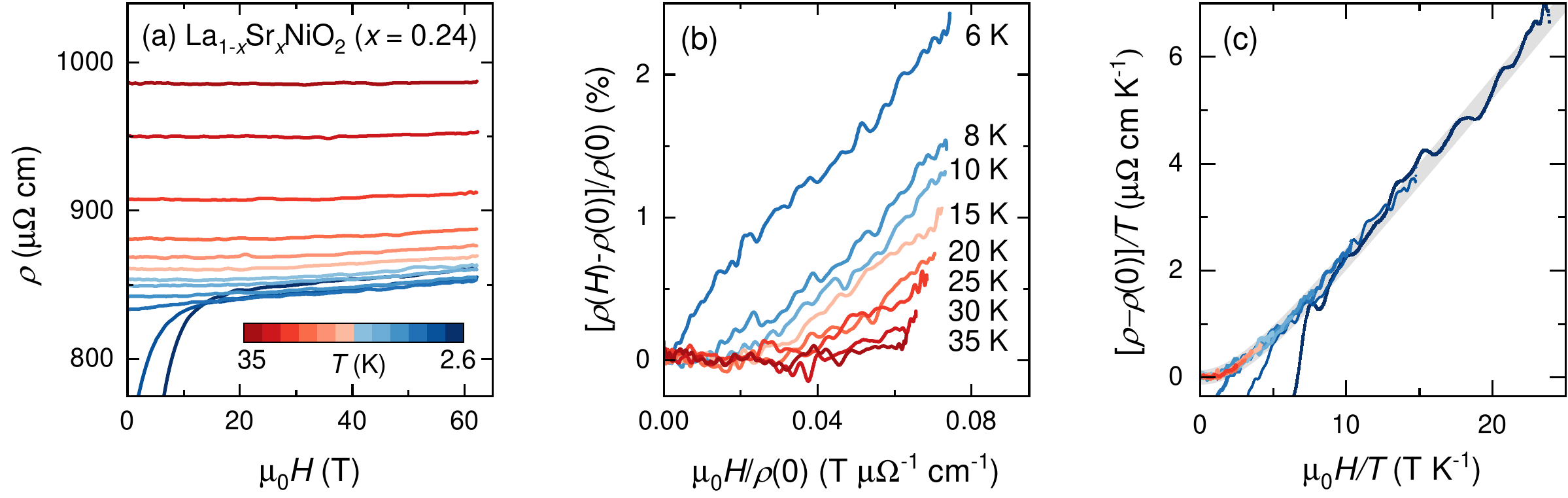}
\caption{
(a) Isothermal magneto-resistivity of overdoped LSNO thin film ($x=0.24$) measured in pulsed magnetic fields up to 62~T at the following temperatures: [35, 30, 25, 20, 17.5, 15, 12.5, 10, 8, 6, 4.2, 2.6] K. Below 7~K, a transition from superconducting to normal state can be clearly seen.
(b) Fractional MR ($\Delta\rho(H)/\rho(0)$) versus magnetic fields divided by zero-field resistivity ($\mu_0H/\rho(0)$), known as the Kohler plot.
(c) Magnetoresistance divided by temperature ($[\rho(H, T)-\rho(0,T)]/T$) versus magnetic fields divided by temperature ($\mu_0H/T$). Gray line is a fit made to the normal-state data using an empirical function: \textcolor{black}{$f(H/T) =\sqrt{1+c(\mu_0H/T)^2}-1$, where $c$ is a numeric constant and $f=0$ at $H=0$ is ensured}.
}
\label{Fig_LSNO24}
\end{figure*}

In the $x=0.20$ film, the doping level closest to $x_{\rm opt}$, the in-plane resistivity $\rho(T)$ exhibits a $T$-linear behavior over an extended temperature range between 300~K and the onset $T_{\rm c}$ (marked by red arrow in Fig.~\ref{Fig_LSNO24}(b)). Similar behavior has been previously found in NSNO and interpreted as the manifestation of strange metallicity \cite{lee2023,iorioDuval2025}. With increasing doping ($x=0.22$ and 0.24), the magnitude of $\rho(T)$ increases and a clear deviation from $T$-linear resistivity is seen, with a positive curvature developed as $T$ decreases. Again, this behavior is consistent with that observed in overdoped NSNO \cite{hsu2024}, pointing to a common evolution of the carrier dynamics with increasing hole doping in the infinite-layer nickelates.
To gain further insights into the normal-state  characteristics in the $T=0$ limit, we use pulsed magnetic fields well above its upper critical field in LSNO and probe its magnetotransport at $T \ll T_{\rm c}$.

\subsection{High-field magnetoresistance}
Figure~\ref{Fig_LSNO24}(a) shows the isothermal magnetoresistivity of the $x=0.24$ sample in magnetic fields up to 62~T at $2.6\leq T\leq 35$~K. The lowest $T_{\rm c}$ in the $x=0.24$ film allows the access of normal-state transport over the widest $(H,T)$ range.
A clear field-induced transition from superconducting to normal- state can be seen below 6~K, with the normal-state MR follows a largely $H$-linear behavior.
Figure~\ref{Fig_LSNO24}(b) show the fractional MR $[\rho(H)-\rho(0)]/\rho(0)$ traces versus $\mu_0H/\rho(0)$, i.e. the Kohler plot, at $T \geq 6$~K. The MR magnitude is small ($\lesssim$2~\%) within this temperature range and practically vanishes at 35~K. The excessive MR increase at 6~K is due to the presence of fluctuating superconductivity at this temperature. 
Evidently, the LSNO ($x=0.24)$ film does not follow Kohler's rule between 8 and 35~K, as also found in overdoped NSNO \cite{hsu2024}.

Notably, the MR functional form evolves from manifestly $H$-quadratic at 30~K to $H$-linear at 6~K, pointing to the unconventional MR form as described by Eq.~(1).
In Figure~\ref{Fig_LSNO24}(c), we plot $[\rho(H,T)-\rho(0,T)]/T$ versus $\mu_0H/T$.
Remarkably, $[\rho(H,T)-\rho(0,T)]/T$ indeed collapse into a single curve when plotted against $\mu_0H/T$, suggesting that the normal-state magnetotransport of overdoped LSNO is not dictated by conventional orbital motion of charge carriers \cite{maksimovic2020,hinlopen2022}.
(Note that for $T\leq T_{\rm c}$, $\rho(0,T)$ is first estimated by fitting the normal-state MR using a linear function: $\rho(H,T)=\rho(0,T)+\gamma'T$, where $\gamma'$ is a $T$-dependent constant; next, we manually adjust $\rho(0,T)$ in $0.1$-$\mu\Omega$-cm steps to achieve the scaling collapse shown in Fig.~\ref{Fig_LSNO24}(c) and registered the corresponding $\rho(0,T)$ as the zero-field normal-state resistivity values.)

\begin{figure}[hbtp!!!]
\includegraphics[width=0.9\linewidth]{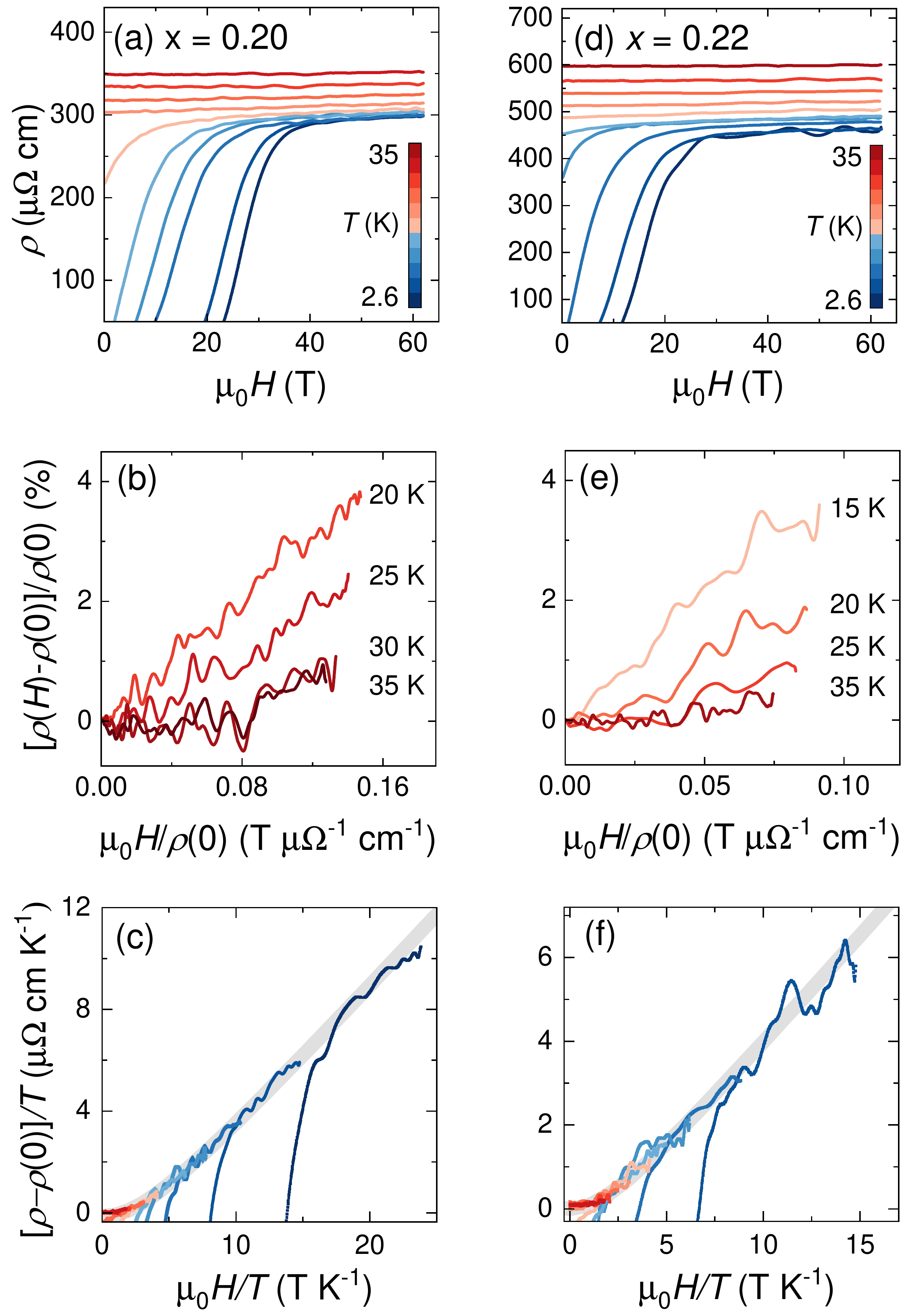}
\caption{(a, d) Isothermal magneto-resistivity of overdoped LSNO films: (a) $x=0.20$ measured at the following temperatures: $T=$ [35, 30, 25, 20, 15, 10, 8, 6, 4.2, 2.6]~K and (d) $x=0.22$ at [35, 30, 25, 20, 15, 12.5, 10, 7, 4.2, 2.6]~K.
(b, e) Kohler plot for (b) $x=0.20$ and (e) $x=0.22$ films.
(c, f) $[\rho(H, T)-\rho(0,T)]/T$ versus $\mu_0H/T$ for (c) $x=0.20$ and (f) $x=0.22$ films. Grey lines are fits made using the empirical function: $\sqrt{1+c(\mu_0H/T)^2}-1$.
}
\label{Fig_LSNO20_22}
\end{figure}

To track the doping evolution of the normal-state MR, we conduct the same analysis on the $x=0.20$ and 0.22 LSNO films as shown in Figure~\ref{Fig_LSNO20_22}.
Although the higher $T_{\rm c}$ of these two films raises the lower bound above which the normal-state MR can be accessed without extrapolations, Kohler's rule is nonetheless clearly violated above $T_{\rm c}$ (Fig.~\ref{Fig_LSNO20_22}(b,e)), in conjunction with the MR functional form evolving from $H^2$ to $H$-linear with decreasing temperature.
Here again, when plotted as $[\rho(H,T)-\rho(0,T)]/T$ versus $\mu_0H/T$, the normal-state MR between 2.6 and 35~K collapse into a single curve within experimental uncertainties (Fig.~\ref{Fig_LSNO20_22}(c,f)) for both $x=0.20$ and 0.22 films. 
Our results demonstrate that the normal-state MR of overdoped LSNO films with $0.20\leq x\leq 0.24$ do not obey Kohler's rule, rather can be described using the MR form as described in Eq.~(1), as found for several putative quantum critical metals \cite{hayes2016,licciardello2019a,giraldogallo2018,sakar2019,ayres2021,ayres2024}. This observation is the first key finding of our study.

\subsection{Normal-state resistivity below \texorpdfstring{$T_{\rm c}$}{Tc}}
\begin{figure*}[hbtp!!!]
\includegraphics[width=0.9\linewidth]{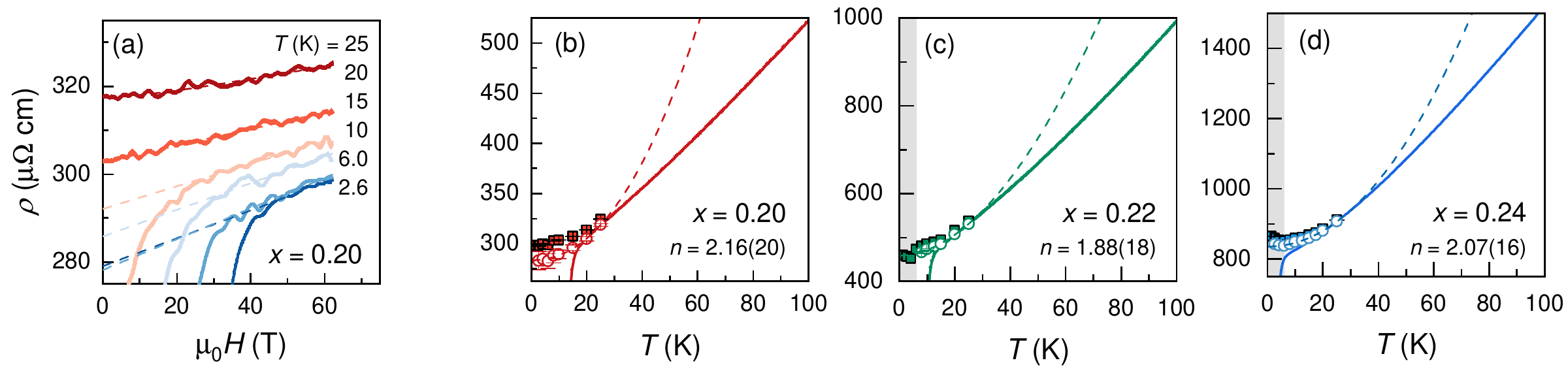}
\caption{
(a) Magnetoresistivity $\rho(H)$ at selected temperatures for $x=0.20$ film. Dashed lines are linear fits made to the normal-state $\rho(H)$ to find initial estimates of the normal-state resistivity at zero field. Open circles on the $H=0$ axis are the corresponding values resulting the $H/T$ scaling collapse in Fig.~\ref{Fig_LSNO20_22}(c), used as the zero-field resistivity shown in panel (b) here.
(b-d) Normal-state resistivity versus temperature for LSNO films with doping as specified. 
Solid lines are measured zero-field resistivity; filled and open points are normal-state resistivity measured at 60~T and extrapolated at 0~T, respectively. Dash lines are fits made to extrapolated $\rho(T, H=0)$ below 25~K using: $\rho(T) = \rho_0 + A_nT^n$. The gray shadings in (c, d) mark the temperature regime in which a slight upturn in $\rho(T)$ is found and excluded from the power-law fits.}
\label{Fig_nsRT}
\end{figure*}

Having established the form of  normal-state MR in overdoped LSNO, next we investigate the temperature dependence of the normal-state resistivity below $T_{\rm c}$. Figure~\ref{Fig_nsRT} shows the normal-state resistivity as a function of temperature below 100~K. For all three overdoped films, both the as-measured resistivity at 60~T and the extrapolated zero-field resistivity show a positive curvature as $T\rightarrow0$, indicating a deviation from putative strange-metal behavior. To extract the resistivity power-law exponent in the $T=0$ limit, we fit the normal-state $\rho(0,T)$ below 25~K using $\rho(T)=\rho(0)+A_nT^n$. As summarized in Table~\ref{table}, the fitted $n$ for all three overdoped films are found to agree with $n_{\rm FL}=2$ expected for a conventional Fermi liquid within experimental uncertainty. The observation of a Fermi-liquid power-law resistivity in the low-temperature normal state of overdoped LSNO is our second key finding. 

\section{Discussion}

\begin{table}[hbtp!!!]
\begin{ruledtabular}
\caption{Extracted normal-state transport parameters for overdoped LSNO films. $n$: normal-state resistivity power-law exponent below 30~K; $\gamma$: MR slope in the high $H/T$ limit (i.e. $\mu_0H/T\gg5$).}
\begin{tabular}{ccccc}
$x$ & $T_{\rm c}$~(K) & $n$ & $\gamma~(\rm\mu\Omega~cm~T^{-1})$\\
\colrule
0.20 & 15.6 & $2.16\pm0.20$ & $0.38\pm0.09$\\
0.22 & 12.6 & $1.88\pm0.18$ & $0.37\pm0.09$\\
0.24 &  6.4 & $2.07\pm0.16$ & $0.34\pm0.08$\\
\end{tabular}
\label{table}
\end{ruledtabular}
\end{table}

Previously, a $H$-linear magnetoresistance has been observed in several unconventional superconductors with $T$-linear resistivity, or the components thereof, which have been interpreted as a property of the strange-metal phase \cite{hayes2016,giraldogallo2018,sakar2019,licciardello2019a,ayres2021,shi2025}. Furthermore, in some cases, $T_{\rm c}$ has been found to correlate closely to the slope of $H$-linear MRf $\gamma$ \cite{ayres2024,shi2025}, hinting at a possible link between the strength of strange metallicity and superconductivity. In overdoped LSNO, however, we find that $\gamma$ is largely independent of nominal hole doping and $T_{\rm c}$ values (see Table~\ref{table}). 
More notably, we find that the $H$-linear MR occurs with a backdrop of $T^2$ resistivity, revealing a new setting for the manifestation of an unconventional MR behavior that follows $H/T$ scaling.

\textcolor{black}{Previous studies \cite{giraldogallo2018,ataei2022,hinlopen2022} have pointed out that it is possible to realize a $H$-linear MR without invoking the anomalous MR form set by Eq.~(1). For example, it was early proposed  that the $H$-linear MR found in La$_{2-x}$Sr$_x$CuO$_4$, could be in principle explained by an elastic scattering rate that is isotropic in $k$-space but magnetic-field dependent \cite{giraldogallo2018}. More recently, numerical calculations based on Boltzmann transport theory have shown that a field-independent but highly $k$-anisotropic elastic scattering rate leads to a $H$-linear MR over an extended field range \cite{ataei2022,hinlopen2022}. Given such calculations require detailed knowledge on the electronic band structure (with the full three-dimensional dispersion parameterized in tight-binding form) and scattering rate anisotropy, which are presently not available, whether the MR behavior in overdoped LSNO can be described using the Boltzmann transport theory remains to be investigated.}

Compared to its Nd-counterpart, whose $\rho(T)$ and $\rho(H)$ can both be described using unconventional power laws with an exponent $\approx$ 1.5 (ref.~\cite{hsu2024}), LSNO exhibits distinct transport behavior in both its $T$- and $H$-dependence. This distinction may be ascribed to the difference in the low-$T$ magnetic ground state of Nd- and La-based infinite-layer nickelates, where NSNO is found to be a spin glass whereas LSNO a short-ranged fluctuating order \cite{fowlie2022}.
\textcolor{black}{Regardless, in both cases, a $T$-linear normal-state resistivity in the $T=0$ limit is not found, which does not support the classification of the infinite-layer nickelates as a standard strange metal in proximity to a QCP.}
Our results highlight the possible impact of rare-earth magnetism on the ground-state characteristics of infinite-layer nickelates \cite{fowlie2022,wang2023,harvey2025}. 
In this regard, the lanthanum-based infinite-layer nickelates may be best positioned to study the physics intrinsic to the NiO$_2$ lattice due to its absence of rare-earth magnetism \cite{osada2021}. 


\section{Summary}
We conduct low-temperature magnetoresistivity on overdoped LSNO thin films with $x = 0.20-0.24$ under pulsed-field conditions up to 62~T. 
In all three films, the magnetoresistivity in the field-induced normal state exhibits a clear violation of Kohler's rule and a $H$-linear behavior in the high $H/T$ limit, reminiscent to that found in several strange metallic systems.
In contrast, the zero-field normal-state resistivity  below 25~K shows a $T^2$ behavior that is characteristic of a Fermi-liquid ground state. 
Our findings demonstrate an intriguing coexistence of strange-metal and Fermi-liquid transport features in overdoped LSNO, and, by comparison with the Nd-counterpart, highlight the impact of rare-earth magnetism to ground-state transport characteristics in infinite-layer nickelates.


\section*{Acknowledgements}
We thank Jake Ayres and Caitlin Duffy for critical review of this manuscript, and the support of Yu-Feng Huang (Instrumentation Center at National Tsing Hua University, Project No. NSTC 115-2740-M-007-001-) for the technical assistance with MPMS3-ETO experiments.
This work is supported by the Yushan Fellow Program (MOE-112-YSFMS-0002-002-P1) and the Center for Quantum Science and Technology (CQST) within the framework of the Higher Education Sprout Project by the Ministry of Education (MOE), Taiwan, and by the National Science and Technology Council, Taiwan (Project No. 113-2112-M-007-044-MY3 and 114-2124-M-007-012-) (Y.C.P., Y.T.H). 
This work is also supported by the Institute for Solid State Physics, The University of Tokyo, grant no. 202403-HMBXX-0047, JSPS KAKENHI, grant nos. JP22K20347 and JP23K13663, the Grant Fund for Research and Education of Institute for Materials Research, Tohoku University; Tohoku University in MEXT Advanced Research Infrastructure for Materials and Nanotechnology in Japan, grant no. JPMXP1224TU0127; the Basic Science Research Projects by The Sumitomo Foundation, the Toyota Riken Scholar Program by Toyota Physical and Chemical Research Institute, the Kazuchika Okura Memorial Foundation, as well as the Thermal and Electric Energy Technology Foundation (M.O.).
We further acknowledge support of the HLD at HZDR, member of the European Magnetic Field Laboratory (EMFL).

\section*{Author contributions}
Y.T.H. conceived the project.
M.O. and A.T. synthesized and characterized the nickelate films.
Y.C.P., T.K., T.H., Y.T.H. performed magneto-resistivity measurements.
Y.C.P. and Y.T.H. analyzed the data and wrote the manuscript with inputs from all authors.

\section*{Competing interests}
The authors declare no competing interests.

\bibliographystyle{apsrev4-1}
\bibliography{reference_nickelates}

\end{document}